\documentclass[10pt]{article}

\usepackage{graphics}
\usepackage{graphicx}

\usepackage[a4paper, left=35mm,right=35mm,top=34mm,bottom=34mm]{geometry}
\usepackage[utf8]{inputenc}
\usepackage[T1]{fontenc}
\usepackage[english]{babel}

\usepackage{enumerate}
\usepackage{graphicx}
\usepackage{hyperref}
\hypersetup{
    colorlinks=true,
    linkcolor=blue,
    filecolor=magenta,      
    urlcolor=cyan,
}
\usepackage{listings}
\usepackage{color}

\definecolor{dkgreen}{rgb}{0,0.6,0}
\definecolor{gray}{rgb}{0.5,0.5,0.5}
\definecolor{mauve}{rgb}{0.58,0,0.82}
\lstdefinelanguage{MRGC++}{%
  language=C++,
  morekeywords={T, U, MPI_Irecv, MPI_Isend, MPI_Allreduce, MPI_Waitall, Compute, Map, abs, max, Swap, MPI_Recv_init, MPI_Send_init, MPI_Startall, Copy, Init, InitRecv, InitSend, InitAllReduce, Send, Recv, AllReduce, Finalize, InitSnapshot, Snapshot, SwitchAsync, SnapReduce, MPI_Test, MPI_Start}
}
\usepackage{mathtools,amsthm,amssymb,amsfonts}
\usepackage{algorithm}
\usepackage{algorithmic}
\makeatother
\theoremstyle{plain}

\theoremstyle{definition}

\theoremstyle{remark}

\usepackage{caption} 
\usepackage{fancyhdr}

\author{
  {\normalsize Guillaume Gbikpi-Benissan}\thanks{IRT SystemX, France.}
  \and
  {\normalsize Fr\'ed\'eric Magoul\`es}\thanks{CentraleSup\'elec, Universit\'e Paris-Saclay, France
    (correspondence, frederic.magoules@hotmail.com).}
}
\title{JACK2: a new high-level communication library for parallel iterative methods}
\date{}

\begin{document}
\maketitle
\thispagestyle{fancy}

\begin{abstract}
\noindent In this paper, we address the problem of designing a distributed application meant to run both classical and asynchronous iterations. MPI libraries are very popular and widely used in the scientific community, however asynchronous iterative methods raise non-negligible difficulties about the efficient management of communication requests and buffers. Moreover, a convergence detection issue is introduced, which requires the implementation of one of the various state-of-the-art termination methods, which are not necessarily highly reliable for most computational environments. We propose here an MPI-based communication library which handles all these issues in a non-intrusive manner, providing a unique interface for implementing both classical and asynchronous iterations. Few details are highlighted about our approach to achieve best communication rates and ensure accurate convergence detection. Experimental results on two supercomputers confirmed the low overhead communication costs introduced, and the effectiveness of our library.
\end{abstract}

\begin{keywords}
parallel computing; iterative methods; asynchronous iterations; convergence detection; message passing interface; application programming interface; parallel programming
\end{keywords}

\section{Introduction}

A wide set of high performance computing (HPC) applications has to face two major challenges consisting of, on one hand, the non-negligible probability of resource failures and, on another hand, their inherent scalability limits due to global synchronization. The scalability of parallel algorithms is experimented by considering the acceleration rate observed when varying either the number of processors, the size of the problem or both. Asynchronous iterations~\cite{ChazMir1969} are gaining more and more attention today, as they remove theoretical upper bounds on maximum expectable acceleration rates, and naturally self-adapt to both unbalanced workload and resource failures. They could thus constitute so far one of the most interesting and simplest approach for HPC software developments, where with such increased number of processors, most part of the computational performance is related to the management of inter-process dependency. Nonetheless, to accurately experiment asynchronous iterative methods, scientists have to deal with quite advanced implementation options, depending on both the computation environment and the communication middleware they use. Few libraries have been proposed, either for Java or C++ applications, namely Java Asynchronous Computing Environment (JACE)~\cite{BahiEtAl2004, BahiEtAl2007} and Communication Routines for Asynchronous Computation (CRAC)~\cite{CoutDom2007}, respectively. Their development was however mainly driven by low-level communication capabilities, especially related to grid environments. They thus do not successfully manage to build upon Message Passing Interface (MPI) libraries, which however are very largely used and maintained in the scientific community. As the MPI framework is designed on a general basis for all kinds of distributed algorithms, particularities of asynchronous iterations are not easy to handle, even less in a non-intrusive manner.

Recently, to the best of our knowledge, Just an Asynchronous Communication Kernel (JACK)~\cite{MagGBen2016} was proposed as the first MPI-based C++ library for parallel implementation of both classical and asynchronous iterations. Main features include (i) a unique high-level application programming interface (API) which allows users to switch to asynchronous iterations at runtime, (ii) a buffer manager which prevents developers from handling memory deallocation for successive outgoing messages, (iii) continuous message reception during processing steps to always deliver last received data, (iv) a convergence detection tool based on the asynchronous iterations termination protocol due to~\cite{BahiEtAl2005}. Despite a lot of efforts to provide a very user-friendly interface, it is still required with JACK to manipulate various objects purely related to the internal behavior of the communication library. Furthermore, it comes out that the convergence detection interface does not suit to other kinds of termination protocol. We therefore present in this paper a new version based on a more flexible and evolutive object-oriented architecture, providing an API with different levels of encapsulation of communication objects.

Section~\ref{sec:fw} first recalls generic parallel schemes of classical (sequential) and asynchronous iterations, respectively, from which implementation issues of the latter are exposed, showing the need for a dedicated library. Section~\ref{sec:ov} presents our approach to implement iterative methods, providing a complete C++ example of a unique JACK-based procedure to run either classical or asynchronous iterations, depending on a user-given argument. Some implementation details are pointed out about the management of communication requests and distributed convergence detection. At last, Section~\ref{sec:ex} reports average results obtained from experiments conducted on supercomputers, for the parallel solution of large sparse linear systems arising from a convection-diffusion problem.

\section{Computational framework}
\label{sec:fw}

\subsection{Classical iterations}

Let $p$ be a strictly positive integer, and let
\[
f : E \to E, \qquad f_{i} : E \to E_{i}, \quad 1 \le i \le p,
\]
be arbitrary mappings, with
\[
E = E_{1} \times \cdots \times E_{p},
\]
and satisfying :
\[
f(x) =
\begin{bmatrix}
f_{1}(x_{1}, \ldots, x_{p}) & \cdots & f_{p}(x_{1}, \ldots, x_{p})
\end{bmatrix}^{\mathsf{T}}, \qquad x \in E.
\]
We consider classical iterations generating a sequence $\{x^{k}\}_{k \in \mathbb{N}}$ such that
\begin{equation}
\label{eq:pia}
x_{i}^{k+1} = f_{i}\left(x_{1}^{k}, \ldots, x_{p}^{k}\right), \qquad 1 \le i \le p.
\end{equation}
Given $p$ processes, Algorithm~\ref{alg:sisc} describes a trivial parallel procedure implementing the computational model~\eqref{eq:pia}.
\begin{algorithm}
\caption{Trivial parallel iterative scheme}
\label{alg:sisc}
\begin{algorithmic}[1]
\STATE{$k := 0$}
\REPEAT
	\STATE{$x_{i}^{k+1} := f_{i}(x_{1}^{k}, \ldots, x_{p}^{k})$}
	\FORALL{$j \in \{1, \ldots, i-1, i+1, \ldots, p\}$}
		\STATE{Request reception of $x_{j}^{k+1}$ from process $j$}
		\STATE{Request sending of $x_{i}^{k+1}$ to process $j$}
	\ENDFOR
	\STATE{Wait for communication completion}
	\STATE{$k := k + 1$}
\UNTIL{$\|x^{k} - x^{k-1}\| \simeq 0$}
\end{algorithmic}
\end{algorithm}
For each iteration on each process, computation is performed first, then messages are sent and received. Any process willing to send data to a slower one therefore needs to wait until the latter finishes its computation phase and requests the reception of the message. With heterogeneous processing and network capabilities, such dedicated communication phases would considerably decrease the parallel computational efficiency.

A well known improvement consists of Algorithm~\ref{alg:siac} which overlaps communication and computation phases.
\begin{algorithm}
\caption{Overlapping parallel iterative scheme}
\label{alg:siac}
\begin{algorithmic}[1]
\STATE{$k := 0$}
\REPEAT
	\FORALL{$j \in \{1, \ldots, i-1, i+1, \ldots, p\}$}
		\STATE{Request reception of $x_{j}^{k+1}$ from process $j$}
	\ENDFOR
	\STATE{$x_{i}^{k+1} := f_{i}(x_{1}^{k}, \ldots, x_{p}^{k})$}
	\FORALL{$j \in \{1, \ldots, i-1, i+1, \ldots, p\}$}
		\STATE{Request sending of $x_{i}^{k+1}$ to process $j$}
	\ENDFOR
	\STATE{Wait for communication completion}
	\STATE{$k := k + 1$}
\UNTIL{$\|x^{k} - x^{k-1}\| \simeq 0$}
\end{algorithmic}
\end{algorithm}
In this scheme, message reception is requested from the beginning of the iteration, therefore a process can start sending data immediately after its computation phase even when the destination process is still computing. This requires a little more memory, since the buffers $x_{j}^{k+1}$ and $x_{j}^{k}$ may be simultaneously accessed, but generally the resolution time is decreased compared to the trivial scheme. Indeed there is almost no more time dedicated to communication on slowest processes, which on another hand does not provide much performance gain when the workload is perfectly balanced.

\subsection{Asynchronous iterations}

Now let $\{P^{k}\}_{k \in \mathbb{N}}$ be a sequence of integer subsets, with
\[
P^{k} \subseteq \{1, \ldots, p\},
\]
and satisfying :
\begin{equation}
\label{eq:ai:ass1}
\forall i \in \{1, \dots, p\}, \quad \operatorname{card}\{k \in \mathbb{N} \ | \ i \in P^{k}\} = \infty.
\end{equation}
Let also $\tau_{j}^{i}$, with $1 \le i \le p$ and $1 \le j \le p$, be nonnegative integer-valued functions, with
\[
\tau_{j}^{i}(k) \le k,
\]
and satisfying :
\begin{equation}
\label{eq:ai:ass2}
\forall i, j \in \{1, \dots, p\}, \quad \lim_{k \to \infty} \tau_{j}^{i}(k) = \infty.
\end{equation}
We consider here asynchronous iterations generating a sequence $\{x^{k}\}_{k \in \mathbb{N}}$ such that
\begin{equation}
\label{eq:apia}
x_{i}^{k+1} = \left \{
\begin{array}{ll}
f_{i}\left(x_{1}^{\tau_{1}^{i}(k)}, \ldots, x_{p}^{\tau_{p}^{i}(k)}\right), & i \in P^{k},\\
x_{i}^{k}, & i \notin P^{k}.
\end{array}
\right.
\end{equation}
The corresponding pattern described by Algorithm~\ref{alg:aiac} is quite simple. As soon as a process terminates an iteration, it starts a next one without waiting for the completion of communication requests. If new data were not received, latest ones are just used in the next computation phase.
\begin{algorithm}
\caption{Asynchronous parallel iterative scheme}
\label{alg:aiac}
\begin{algorithmic}[1]
\STATE{$k_{i} := 0$}
\REPEAT
	\FORALL{$j \in \{1, \ldots, i-1, i+1, \ldots, p\}$}
		\STATE{Request reception of $x_{j}^{\tau_{j}^{i}(k)}$ from process $j$}
	\ENDFOR
	\STATE{$x_{i}^{k_{i}+1} := f_{i}(x_{1}^{\tau_{1}^{i}(k)}, \ldots, x_{i}^{k_{i}}, \ldots, x_{p}^{\tau_{p}^{i}(k)})$}
	\FORALL{$j \in \{1, \ldots, i-1, i+1, \ldots, p\}$}
		\STATE{Request sending of $x_{i}^{k_{i}+1}$ to process $j$}
	\ENDFOR
	\STATE{$k_{i} := k_{i} + 1$}
\UNTIL{$\|f(x_{1}^{k_{1}}, \ldots, x_{p}^{k_{p}}) - (x_{1}^{k_{1}}, \ldots, x_{p}^{k_{p}})\| \simeq 0$}
\end{algorithmic}
\end{algorithm}

Implementing such an algorithmic scheme obviously raises nontrivial questions related to the management of the successive communication requests, the management of associated buffers and the evaluation of the stopping criterion. On performance side for instance, it might be preferable to allow for several message reception requests during computation phase, in order to use the least delayed data. On another hand, one notices that overlapping and asynchronous schemes nearly follow the same algorithmic pattern. It is therefore desirable to also be able to provide a unique implementation for both, and just consider a runtime parameter to produce either classical or asynchronous iterations. Next section presents main features of the API of our MPI-based communication library which handles all these implementation issues.

\section{Implementing iterative methods}
\label{sec:ov}

\subsection{Preprocessing steps}

Full initialization of a JACK2 communicator object is the sole actual delicate part while using the library, as it requires to clearly identify and prepare variables for communication graph, communication buffers, computation residual and solution vectors. Such an approach rigorously follows from the computational models previously described but of course, our design is flexible enough to provide several ways for interacting with the library. The communication graph is generally distributed upon the set of processes such that each process handles the list of its one-hop neighbors in the graph. We explicitly distinguish outgoing and incoming communication links, as shown by Listing~\ref{lst:cg}.
\begin{lstlisting}[caption=Communication graph, label=lst:cg]
/* template <typename T, typename U> */
// T: float, double, ...
// U: int, long, ...
U* sneighb_rank; // MPI rank of neighbors on outgoing links.
U* rneighb_rank; // MPI rank of neighbors on incoming links.
U numb_sneighb;
U numb_rneighb;
\end{lstlisting}
Accordingly, for each communication link, an array buffer is accessed to either write or read computation data. One thus need, for each buffer, its size and memory address, as shown by Listing~\ref{lst:cb}.
\begin{lstlisting}[caption=Communication buffers, label=lst:cb]
/* template <typename T, typename U> */
// T: float, double, ...
// U: int, long, ...
T** send_buf; // buffers for sending computation data.
T** recv_buf; // buffers for receiving computation data.
U* sbuf_size;
U* rbuf_size;
\end{lstlisting}

To terminate the iterations loop, one has to evaluate a stopping criterion depending on a norm-based distance between two potential solutions. In synchronous computation, a distributed residual vector is easily deduced and then its norm is obtained by performing an MPI reduction operation. Each process handles a block-component of the residual vector, which we shall denote by the local residual vector. Whole distributed vectors will then be designated sometimes as global. Mostly, distance functions are based on classical vector norms given by:
\[
\|x\|_{q} = \left(\sum_{i}|x_{i}|^{q}\right)^{1/q}, \qquad q \ge 1.
\]
In Listing~\ref{lst:cr}, the value of $q$ is considered as the type of the norm, and setting $q < 1$ will designate the use of the maximum norm
\[
\|x\|_{\infty} = \max_{i}|x_{i}|.
\]
\begin{lstlisting}[caption=Computation residual, label=lst:cr]
/* template <typename T, typename U> */
// T: float, double, ...
// U: int, long, ...
T* res_vec_buf; // local residual vector.
U res_vec_size;
T res_vec_norm; // norm of the global residual vector.
float norm_type; // 2 for Euclidean norm, < 1 for maximum norm.
\end{lstlisting}
Finally, the stopping criterion
\[
\|f(x_{1}^{k_{1}}, \ldots, x_{p}^{k_{p}}) - (x_{1}^{k_{1}}, \ldots, x_{p}^{k_{p}})\| \simeq 0
\]
in Algorithm~\ref{alg:aiac} clearly points out a convergence detection problem under asynchronous iterations. One has to isolate a unique distributed vector
\[
\begin{bmatrix}
x_{1}^{k_{1}} & \cdots & x_{p}^{k_{p}}
\end{bmatrix}^{\mathsf T},
\]
to which the iterations mapping $f$ must then be applied, all of this in a distributed non-blocking manner. First methods for terminating asynchronous iterations imply a modification of the algorithmic pattern, which also impact the computational model~\eqref{eq:ai:ass1}--\eqref{eq:apia} (see, e.g.,~\cite{BertTsit1991, ElBaz1996}). For being less intrusive, and then more widely applicable, supervised termination should be considered instead, where methods are designed to deduce global convergence from monitoring the local convergence of each process (see, e.g.,~\cite{SavBert1996, BahiEtAl2008}). Yet for now, the snapshot-based approach due to~\cite{SavBert1996} provides the sole practical termination algorithms which allow us to build and evaluate a global solution vector. Listing~\ref{lst:cs} specifies involved variables.
\begin{lstlisting}[caption=Computation solution, label=lst:cs]
/* template <typename T, typename U> */
// T: float, double, ...
// U: int, long, ...
T* sol_vec_buf; // local solution vector.
U sol_vec_size;
int lconv_flag; // armed to notify local convergence state.
\end{lstlisting}

\subsection{Overall features}

\begin{figure}
\begin{center}
\includegraphics[scale=0.75]{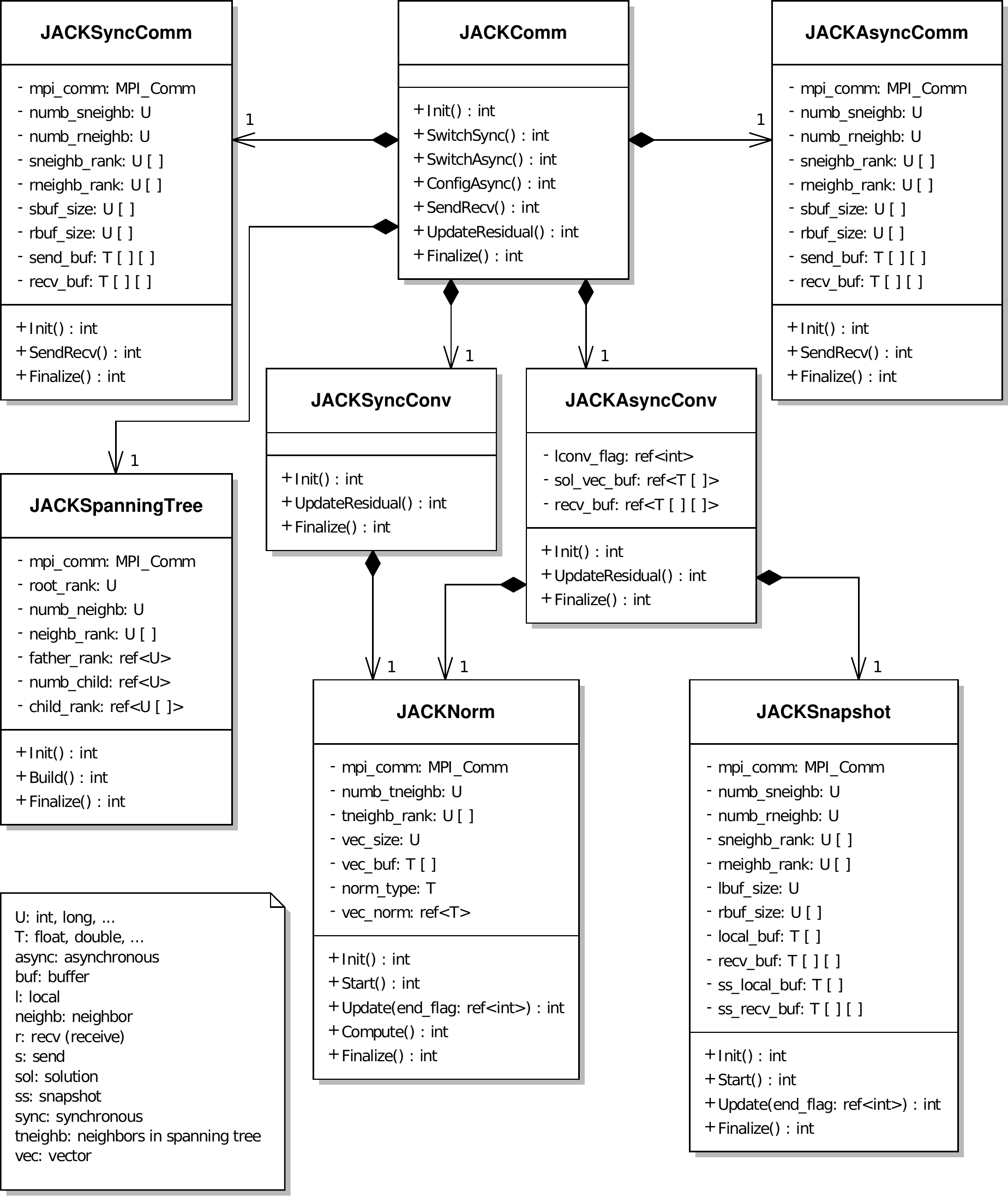}
\caption{Main classes with basic features.}
\label{fig:jack:api}
\end{center}
\end{figure}

Summarily, the design of the library is driven by two general needs. First, processes request an exchange of computation data at each iteration, based on a graph of communication (logical network). We provide classes \emph{JACKSyncComm} and \emph{JACKAsyncComm} with a unique interface to perform this task in a blocking and a nonblocking way, respectively. For asynchronous iterations, the class \emph{JACKAsyncComm} automatically ensures multiple message receptions during computation phases. The second need is the ability to define and evaluate a stopping criterion of the iterations loop. To this end, classes \emph{JACKSyncConv} and \emph{JACKAsyncConv} evaluate a norm-based residual error corresponding to the iterated vector solution. They thus depend on a class \emph{JACKNorm} which computes the norm of any given distributed vector, by using a leader election protocol designed for acyclic graphs. This finally implies a class \emph{JACKSpanningTree} to build a distributed spanning tree over the initial logical network. Under asynchronous iterations, identifying an iterated vector is not straightforward, therefore a class \emph{JACKSnapshot} is used by \emph{JACKAsyncConv} to assemble a consistent distributed vector from the independently iterated block-components. At last for a unique implementation of both classical and asynchronous iterations, a class \emph{JACKComm} can be the only used front-end tool which provides both the data exchange and the convergence detection interfaces. Figure~\ref{fig:jack:api} gives an overview of the architecture of the library, with the basic interface of each class.

JACK2 has been thought to reduce MPI-aware application development to its minimal requirements. First, one has to initialize a communicator object by providing information about the communication graph, communication buffers, computation residual and solution vectors (see Listing~\ref{lst:cg} to Listing~\ref{lst:cs}). Listing~\ref{lst:init} displays an example of initialization part.
\begin{lstlisting}[caption=Initialization, label=lst:init]
// -- initializes MPI
MPI_Init(&argc, &argv);

// -- initializes JACK2 communicator
JACKComm comm;
comm.Init(numb_sneighb, numb_rneighb, sneighb_rank, rneighb_rank, MPI_COMM_WORLD);
comm.Init(sbuf_size, rbuf_size, send_buf, recv_buf);
comm.Init(res_vec_size, res_vec_buf, &res_vec_norm, norm_type);

// -- initializes asynchronous mode
if (async_flag) {
	comm.ConfigAsync(sol_vec_size, &sol_vec_buf, &lconv_flag, &recv_buf);
	comm.SwitchAsync();
}
\end{lstlisting}
The norm of the global residual vector is computed by the library, it is therefore the memory address of the corresponding variable which must be provided (output parameter). The flag indicating the local convergence state of a process may be armed or disarmed at any iteration. As it constitutes an input parameter for the internal behavior of the library, this requires to continuously update the communicator object. Instead of repeated transfers of the state of the flag, one can provide its memory address once for all. We previously saw that snapshot-based termination of asynchronous iterations isolates a global solution vector from which a global residual is deduced. Global iterated solution vectors are distributed to each process through local and received block-components. The corresponding variables can therefore be used as outputs of the JACK2 snapshot protocol, in order to iterate on the isolated global solution without any other particular computation. A global residual evaluation would thus follow in an unnoticed non-intrusive manner. This implied however to provide memory addresses of the variables. Processing steps are thus made very simple, as shown by Listing~\ref{lst:iter}. Implementation details are discussed in the next section.
\begin{lstlisting}[caption=Iterations, label=lst:iter]
comm.Send();
while (res_vec_norm >= 1e-8) {
	comm.Recv();
	// computation phase input: recv_buf, sol_vec_buf.
	// computation phase output: send_buf, sol_vec_buf, res_vec_buf.
	Compute(recv_buf, sol_vec_buf, send_buf, sol_vec_buf, res_vec_buf);
	comm.Send();
	lconv_flag = (Norm(res_vec_buf) < 1e-8);
	comm.UpdateResidual();
}
\end{lstlisting}

\subsection{Communication requests}

As soon as a JACK2 communicator object is initialized with communication graph and buffers, incoming communication channels are opened by MPI non-blocking reception requests, so that each process is ready to receive computation data from any of its neighbors. In synchronous mode, the \emph{Recv} method of \emph{JACKComm} is therefore meant to deliver pending messages from all of the neighbors, and does not return until then. This implies the use of temporary reception buffers, however message delivering is quickly performed by exchanging memory addresses instead of copying whole buffers (see Algorithm~\ref{alg:recv}).
\begin{algorithm}
\caption{Message reception in synchronous mode}
\label{alg:recv}
\begin{algorithmic}[1]
	\FORALL{i := 1..numb\_rneighb}
		\STATE{Wait for the completion of the active MPI reception request}
		\STATE{Exchange memory addresses of the temporary and the user reception buffers}
		\STATE{Activate an MPI reception request on the temporary buffer}
	\ENDFOR
\end{algorithmic}
\end{algorithm}

For asynchronous iterations, incoming messages should be directly received without the need of an explicit request from the user application. Here therefore, the \emph{Recv} method only ensures that incoming channels stay opened after message reception. Indeed, the computational performance of asynchronous iterations relies in part on the ability for each process to receive several messages during computation, so that it benefits from the least delayed data. With the first JACK library, this is managed by the member function \emph{StartContinuousRecv} of the class \emph{Jack::Communicator}, which makes use of multi-threading capabilities. Here, JACK2 rather estimates, for each incoming communication channel, an appropriate number of MPI reception requests that should be active during computation phases (see Algorithm~\ref{alg:irecv}). Still, users now have the possibility to configure this maximum message reception rate.
\begin{algorithm}
\caption{Message reception in asynchronous mode}
\label{alg:irecv}
\begin{algorithmic}[1]
\FORALL{i := 1..numb\_rneighb}
	\FORALL{j := 1..max\_numb\_request}
		\IF{MPI reception request is completed}
			\STATE{Activate an MPI reception request on the user buffer}
		\ENDIF
	\ENDFOR
\ENDFOR
\end{algorithmic}
\end{algorithm}
This implies on the other hand that allowing a higher message sending rate could lead to a counter performance, as the number of pending MPI sending requests may quickly increase, which would yield much more delayed iterations data. JACK2 sending requests are therefore discarded for busy outgoing communication channels (see Algorithm~\ref{alg:isend}).
\begin{algorithm}
\caption{Message sending in asynchronous mode}
\label{alg:isend}
\begin{algorithmic}[1]
\FORALL{i := 1..numb\_sneighb}
	\IF{MPI sending request is completed}
		\STATE{Activate an MPI sending request on the user buffer}
	\ENDIF
\ENDFOR
\end{algorithmic}
\end{algorithm}

\subsection{Convergence detection}

We shall finally focus on the asynchronous convergence detection problem, which constitutes the very main motivation for a communication library designed for massively parallel iterative computing. JACK2 relies on the snapshot-based approach from~\cite{SavBert1996}, where several network configurations are considered. The most decentralized protocol therein consists of, (i) a coordination phase designed upon a spanning tree of the original communication graph, (ii) followed by the actual snapshot phase which is directly performed on the original graph. In the coordination phase, local convergence is notified by processes from the leaves to the root of the spanning tree. When local convergence is observed on a leaf process, a notification is sent to its father in the tree. Internal node processes do the same when, additionally, all of their children have notified local convergence. Under same conditions, the root process instead triggers the snapshot phase as described by Algorithm~\ref{alg:sb96_init}.
\begin{algorithm}
\caption{Snapshot phase initiation for convergence detection (root process)}
\label{alg:sb96_init}
\begin{algorithmic}[1]
\STATE{ss\_sol\_vec\_buf := sol\_vec\_buf}
\STATE{ss\_send\_buf := send\_buf}
\FORALL{i := 1..numb\_sneighb}
	\STATE {Send snapshot message ss\_send\_buf[i] to process sneighb\_rank[i]}
\ENDFOR
\end{algorithmic}
\end{algorithm}
Processes then follow rules given by Algorithm~\ref{alg:sb96_ss_sol} and Algorithm~\ref{alg:sb96_ss_recv}, to isolate a global solution vector distributed on each process as local and received block-components. By finally exchanging memory addresses, the next iteration of each process will be automatically performed on this global vector, so that a consistent global residual will also be evaluated.
\begin{algorithm}
\caption{Snapshot of local solution vector (non-root processes)}
\label{alg:sb96_ss_sol}
\begin{algorithmic}[1]
\IF {lconv\_flag = 1 and at least one snapshot message received}
	\STATE{ss\_sol\_vec\_buf := sol\_vec\_buf}
	\STATE{ss\_send\_buf := send\_buf}
	\FORALL{i := 1..numb\_sneighb}
		\STATE {Send snapshot message ss\_send\_buf[i] to process sneighb\_rank[i]}
	\ENDFOR
\ENDIF
\end{algorithmic}
\end{algorithm}
\begin{algorithm}
\caption{Snapshot of reception buffers}
\label{alg:sb96_ss_recv}
\begin{algorithmic}[1]
\IF {snapshot message ss\_data received from a process rneighb\_rank[j]}
	\STATE{ss\_recv\_buf[j] := ss\_data}
\ENDIF
\end{algorithmic}
\end{algorithm}

\section{Numerical experiments}
\label{sec:ex}

\subsection{Problem and experimental settings}

\begin{figure}
\begin{center}
\includegraphics[scale=0.12]{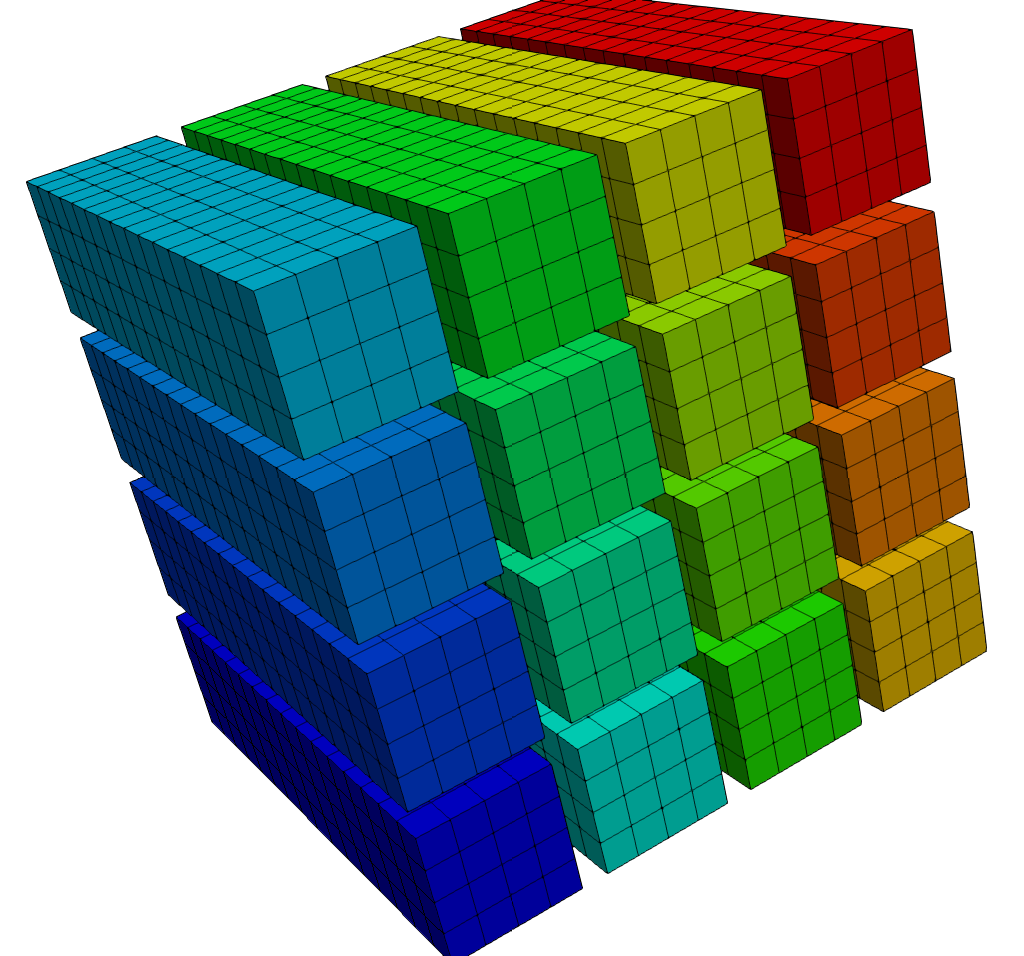}
\caption{Domain discretization and partitioning (e.g., 16 sub-domains).}
\label{fig:partition}
\end{center}
\end{figure}

We consider the experimentation of JACK2 for the solution of a convection-diffusion problem
\[
\frac{\partial u}{\partial t} - \nu \Delta u + a.\nabla u = s, \qquad t \in \mathbb{R}^{+},
\]
where $u$ and $s$ are functions defined on $\mathbb{R}^{+} \times ([0, 1])^{3}$. Conditions and parameters are set to arbitrary values
\[
\left\{
\begin{array}{lll}
u(0,x,y,z) & = & 0, \quad \forall x, y, z \in (0, 1),\\
u(t,x,y,z) & = & 0, \quad \forall x, y, z \in \{0, 1\}, \forall t \in \mathbb{R}^{+},\\
\nu & = & 0.5,\\
a & = & (0.1, -0.2, 0.3).
\end{array}
\right.
\]
By using a finite-difference discretization and the backward Euler integration scheme, we obtain sparse linear systems
\[
\mathcal A U^{t_{n}} = B^{t_{n}, t_{n-1}}, \qquad t_{n} \in \mathbb{R}^{+}, \quad n \in \mathbb{N}^{*},
\]
with $U^{t_{n}}, B^{t_{n}, t_{n-1}} \in \mathbb R^{m}$, $m \in \mathbb N$ and $t_{0} = 0$, for which we find approximated solutions $\widetilde U^{t_{n}}$ by means of both Jacobi and asynchronous relaxations. Figure~\ref{fig:partition} illustrates the geometrical discretization and decomposition of the domain $([0, 1])^{3}$.
Each process handles exactly one sub-domain, and the number of processes always equals the number of processor cores used. Most of the simulations have been run for 5 time steps of size $\delta t = 0.01$.

Experiments on less than 512 processor cores are led on a cluster of 68 nodes Altix ICE 8400 LX with QDR Infiniband interconnect (40 Gb/s). Each node consists of two 6-cores Intel Xeon X5650 CPU at 2.66 GHz, and 21 GB RAM allocated to parallel jobs. The SGI-MPT middleware was loaded as MPI library. For 512 or higher numbers of cores, a cluster of 5040 nodes Bullx B510, also with QDR Infiniband interconnect, has been targeted. Each node there consists of two 8-cores Intel Sandy Bridge E5-2680 CPUs at 2.7 GHz, and 64 GB RAM. The Bullxmpi middleware is used as MPI library.

\subsection{Numerical results}

\begin{figure}
\begin{center}
\includegraphics[scale=0.12]{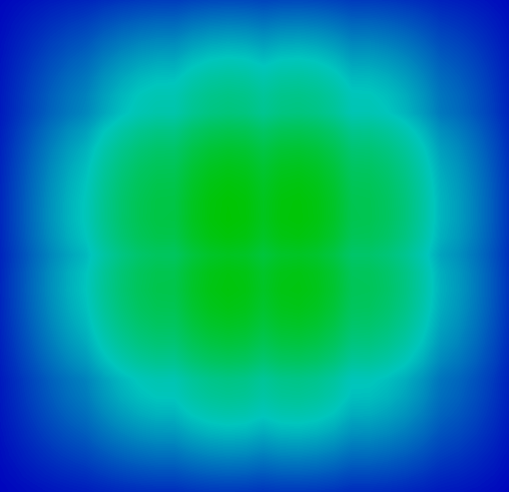}
\includegraphics[scale=0.12]{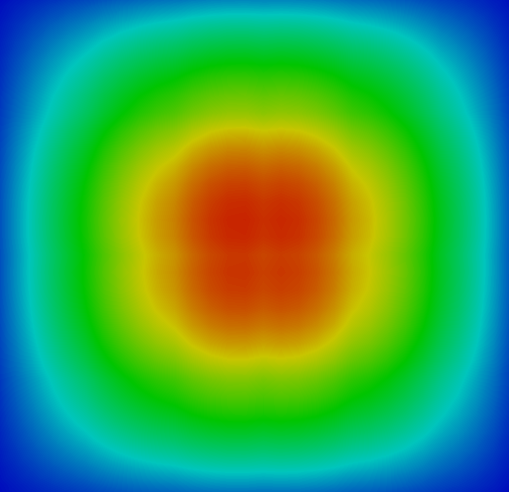}
\includegraphics[scale=0.12]{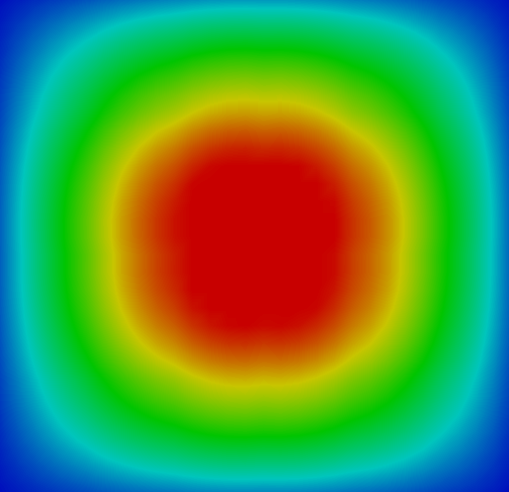}
\includegraphics[scale=0.12]{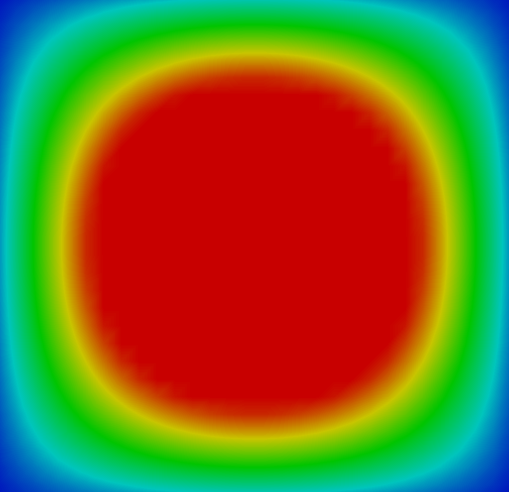}\\
\ \\
\includegraphics[scale=0.12]{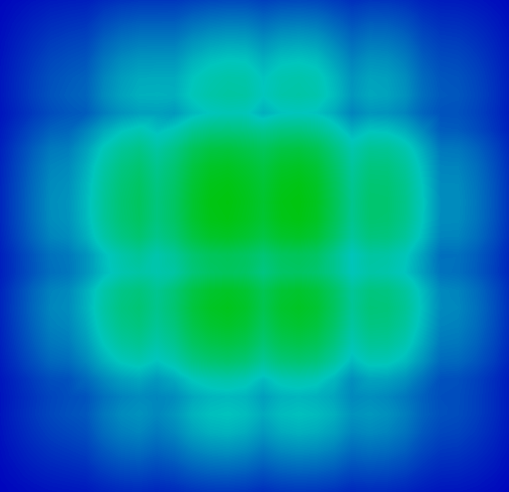}
\includegraphics[scale=0.12]{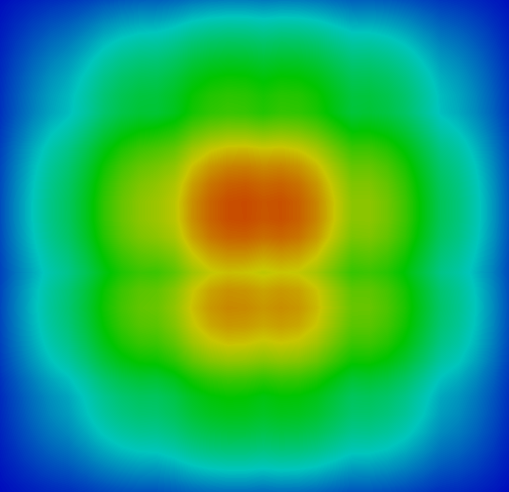}
\includegraphics[scale=0.12]{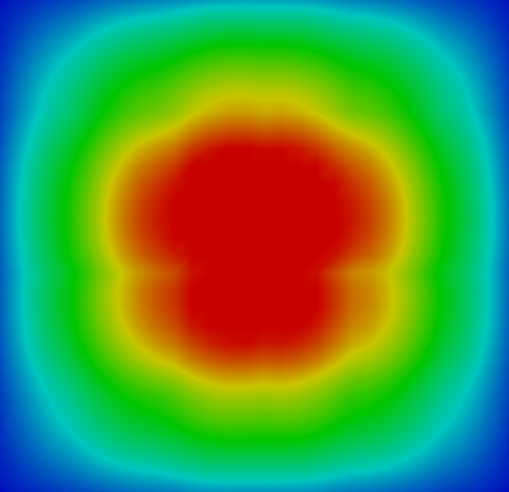}
\includegraphics[scale=0.12]{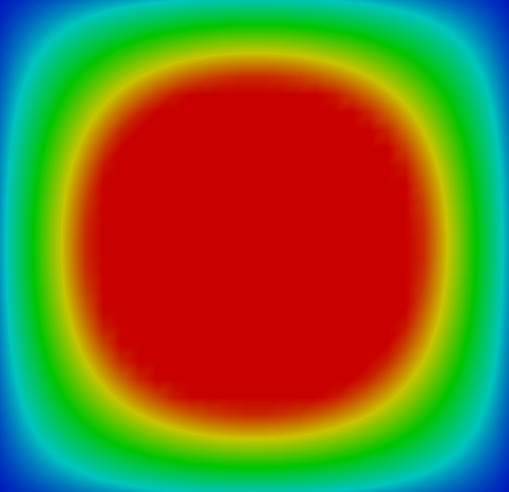}
\caption{Comparison example of classical (top) and asynchronous (down) iterations.}
\label{fig:iterations}
\end{center}
\end{figure}

We see on Figure~\ref{fig:iterations} an example of iterations behavior, both classical and asynchronous. One can effectively notice, for asynchronous iterations, the discontinuity of the iterated solution over the interface between the sub-domains (16 in this example), however convergence is eventually reached compared to the classically iterated solution.

Table~\ref{tab:res} reports some average experimental results from several executions of both the Jacobi and the asynchronous iterative methods. It features, (i) $p$, the number of sub-domains (or processor cores), (ii) $\sqrt[3]{m}$, the cube root of the problem size, (iii) the execution time of one time step resolution, (iv) $r_{n}$, the final residual error observed after termination, (v) and either the corresponding number of Jacobi iterations or the number of snapshots executed during asynchronous iterations. We conclusively observe a significant performance gain from asynchronous iterations, especially on the Bullx B510 cluster ($p \ge 512$), which seems to be consistent with the higher termination delay occurring on the Altix ICE cluster. One should also point out the low communication overhead cost introduced by our implementation of the convergence detection method, since a higher number of snapshots tends to improve the termination delay.
\begin{table}[!ht]
\begin{center}
\begin{tabular}{|c|c|c|}
\hline
& Jacobi relax. & Asynchronous relax.\\
\hline
\begin{tabular}{c|c}
$p$ & $\sqrt[3]{m}$\\
\hline
120 & 180\\
240 & 180\\
420 & 180\\
512 & 175\\
1024 & 180\\
2048 & 185\\
4096 & 188
\end{tabular}
&
\begin{tabular}{c|c|c}
Time & $r_{n}$ & \# Iter.\\
\hline
490 & 8.3e-7 & 127081\\
281 & 8.3e-7 & 129031\\
183 & 8.3e-7 & 131046\\
36 & 8.3e-7 & 80611\\
50 & 8.3e-7 & 135595\\
90 & 8.3e-7 & 312520\\
226 & 8.3e-7 & 736287
\end{tabular}
&
\begin{tabular}{c|c|c}
Time & $r_{n}$ & \# Snaps.\\
\hline
491 & 5.5e-7 & 9\\
250 & 2.5e-7 & 20\\
154 & 3.5e-7 & 7\\
26 & 9.1e-7 & 20\\
26 & 7.0e-7 & 24\\
39 & 7.7e-7 & 46\\
57 & 6.7e-7 & 90
\end{tabular}\\
\hline
\end{tabular}
\hfill\\
\hfill\\
$r_{n} = \| \mathcal A \widetilde U^{t_{n}} - B^{t_{n}, t_{n-1}} \|_{\infty}$, $\quad \widetilde U^{t_{n}}, B^{t_{n}, t_{n-1}} \in \mathbb R^{m}$.\\
\# Iter. : number of iterations.\\
\# Snaps. : number of snapshots.
\caption{Average results with JACK2-based implementation. Execution time in seconds and residual threshold set to 1e-6.}
\label{tab:res}
\end{center}
\end{table}

\section{Conclusion}

Building a distributed application upon the standard MPI specification to experiment asynchronous iterations is not a straightforward implementation task. Among the software libraries for asynchronous iterations that have been published, namely JACE, JACEV, CRAC and JACK, only the latter is based on MPI. This paper briefly presented a more flexible and evolutive design of such an MPI-based library, which proposes:
\begin{itemize}
\item a new API with different levels of encapsulation of communication objects,
\item a new convergence detection tool based on~\cite{SavBert1996}, successfully encapsulated in a non-intrusive manner,
\item the possibility now to add various other termination protocols,
\item distributed non-blocking computation of vector norms, which will easily evolve to integrate MPI 3 non-blocking collective routines,
\item a simplified management of continuous message reception, with a configurable behavior,
\item and tunable features for advanced experiments.
\end{itemize}
Above all, JACK2 is thus the first library which can be used to compute a global residual error under asynchronous iterations. Extensive experiments on supercomputers with up to 4096 processor cores have been successfully conducted for the solution of large sparse linear systems arising from the discretization of partial differential equations.

\bibliography{ref}
\bibliographystyle{abbrv}

\end{document}